\documentclass[aps,prl,twocolumn,showpacs,groupedaddress,amsmath,amssymb]{revtex4}
\usepackage[dvips,usenames]{color}
\usepackage{nicefrac}		
\usepackage{bbm} 				
\usepackage{graphicx}
\usepackage{hyperref}
\usepackage[all]{hypcap}
\usepackage{leftidx}		

\def \be{\begin{displaymath}}
\def \ee{\end{displaymath}}              

\def \ben{ \begin{equation} }
\def \een{ \end{equation}   }            

\def \bea{\begin{eqnarray*}}             
\def \eea{\end{eqnarray*}}

\def \bean{\begin{eqnarray}}             
\def \eean{\end{eqnarray}}

\def \nn{\nonumber}

\def \cl#1{ {\cal #1} }               
\def \b#1{ \mathbf{#1} }              
\def \tr{ \mbox{{\rm tr}}}
\def \sp#1{ \langle #1 \rangle }      

\def \Ref#1{(\ref{#1})}

\def \inv{ ^{-1} }
\def \dag{\;\!\!^\dagger}

\def \fr#1#2{ \frac{#1}{#2} }

\pacs{03.67.Bg, 02.50.Ga, 03.65.Yz, 03.67.Mn}

\begin{document}

\title{Distance dependence of entanglement generation via a bosonic heat bath}
\author{Thomas Zell, Friedemann Queisser, and Rochus Klesse}
\affiliation{Universit\"at zu K\"oln, Institut f\"ur Theoretische
  Physik, Z\"ulpicher Str. 77, D-50937 K\"oln, Germany} 
\date{March 18, 2009}
\begin{abstract}
Within a generalized Caldeira-Leggett model we analyze
the conditions under which a bosonic heat bath can entangle two 
microscopic quantum systems at a distance $r$. 
We find that the attainable entanglement is extremely distance-sensitive. 
Significant entanglement can only be achieved if the systems are
within a {\em microscopic} distance that is
of order of the cut-off wavelength $\lambda$ of the 
system-bath interaction. At larger distances the maximal entanglement
is exponentially suppressed with a decay length of order $\lambda$. 
We conclude that entanglement generation via a heat bath is not
suitable for entangling remote objects. 
\end{abstract}

\maketitle

Establishing and preserving quantum-mechanical entanglement
\cite{HHHH07} between two remote microscopic physical systems is an experimentally
challenging undertaking. Generally speaking, the difficulties arise
from the fact that the systems need to significantly interact with
each other (in order to build up an entangled common state), and at
the same time the two systems must be thoroughly shielded from
interaction with external degrees of freedom (in order to preserve the
entangled state against decoherence\cite{JZKGKS03}). 
In most physical situations these two requirements appear to be
contradicting to each other and therefore can be only partially fulfilled.

During the last years a fresh look at this entangling
dilemma has emerged from theoretical work on the dynamics of entanglement
in open systems, notably from the work of Braun \cite{Bra02,Bra05} and Benatti
et al. \cite{BFP03,BF06}. It has been shown that
under suitable conditions two two-level systems 
\cite{Bra02,Bra05,BFP03,OK06,STP06} or two harmonic oscillators 
\cite{BF06,ho-work} can become entangled
by mere interaction with a common bosonic heat bath, without
any direct interaction between the microscopic systems. 
In such a situation the coupling to the heat bath has two relevant effects:
it leads to decoherence, as it usually does, but it also mediates an
effective interaction between the systems. When the latter one is strong
enough to overcompensate the decohering effect, the coupling to the
heat bath may eventually lead to entangled microscopic systems. 
Entanglement generation via environmental
modes is, needless to say, a sophisticated  mechanism. Its theoretical
analysis therefore necessarily 
has to rely on idealizing assumptions, and it is still not clear to which 
extent these assumptions can be met in real systems.

We pursued research that especially addresses the role of the spatial
distance $r$ between the microscopic systems on the entangling
mechanism. In doing so,  we {\em fully} account for {\em dissipative}
system-bath interactions by investigating an exactly solvable model
along the lines of Ullersma \cite{Ull66}, and Caldeira and Leggett 
\cite{CL83}. Existing studies \cite{Bra05,STP06} of the distance dependence are either
confined to a {\em dissipation-free} spin-boson model or treat dissipation
on a perturbative level only. 
In contrast to the comparatively moderate power-law dependence observed in 
\cite{Bra05,STP06}, here we find
significant entanglement between 
the systems  only if the distance $r$ does not much exceed the cut-off
wavelength $\lambda$ of the system-bath interaction. 
At larger distances, $E_{max}$, the maximum attainable logarithmic negativity
(as a measure of entanglement \cite{VW02}), decreases exponentially
with a decay length of order $\lambda$. 
We argue that $\lambda$ will be typically of order
of the spatial extension of the microscopic systems and thus conclude 
that entanglement generation via a heat bath is limited to truly
microscopic distances only. 

In our model, the two remote microscopic quantum systems
are represented by two identical harmonic oscillators located on a line at positions
$x_{1/2}=\pm r/2$. 
The choice of harmonic oscillators makes the model exactly solvable. 
Nevertheless, we still expect it to capture the basic 
physical behaviour of any system with a discrete spetrum
(cf.~\cite{SH04}). 
The oscillators, henceforth called the {\it system
oscillators}, have mass $m$ and frequency $\omega_0$, and $P_1,Q_1$
and $P_2,Q_2$ denote their canonical variables. 
They are coupled to an extended, one-dimensional heat bath
consisting of symmetric ($\propto\cos kx$) and antisymmetric
($\propto\sin kx$) harmonic modes of wavenumbers $k>0$ and
frequencies $\omega_k = c k$, $c$ being 
the velocity of sound/light. Let $p_k^{s/a}, q_k^{s/a}$ denote their
respective canonical variables.
We consider a bilinear system-bath interaction $H_I$, where the two
oscillators locally couple to symmetric and antisymmetric bath-modes in the 
same manner,
\be
H_I =  \sum_k g_k (Q_1+Q_2) q_k^s \cos \frac{k r}{2} 
 + g_k (Q_1-Q_2) q_k^a \sin \frac{k r}{2} \:.
\ee
The coupling strengths $g_k$ may be characterized as usual by a
spectral function $J(\omega) := \sum_k \fr{g_k^2}{2 m_k \omega_k}
\delta(\omega - \omega_k)$. 
Here, we assume $J(\omega)$ to be linear for small $\omega$ with a
Drude cut-off,
$
J(\omega) = \fr{2 m \gamma}{\pi} \: \omega \: \fr{\Omega^2}{\Omega^2 +
  \omega^2}\:,
$
leading to ohmic damping with a damping constant $\gamma$.
Typically, the cut-off frequency $\Omega \equiv 2\pi c/\lambda$
is {\em not} some intrinsic frequency of the
bath, rather, it will be determined by the physics
of the system-bath coupling.  
Then, since generally $|g_k|$ markedly declines when $|k|\inv$ falls below
the spatial extension $l$ of the microscopic systems, a good order of
magnitude estimate is $\lambda \sim l$, 
meaning that $\Omega \sim 2 \pi c/ l$.

We also include a counter-term $V_c = \sum_k \frac{g_k^2}{2 m_k
  \omega_k^2}(Q_1^2 + 2 Q_1 Q_2 \cos(k r) + Q_2^2 )$ in the
total Hamiltonian. Its purpose is twofold: firstly, it removes the frequency
renormalization caused by the coupling to the bath \cite{CL83}, secondly, it ensures
that the Quantum Langevin Equations (QLE) which we are going to derive
below will only contain retarded couplings between the oscillators.

The dynamics of the system oscillators can be approached
by means of the Heisenberg equations of motions for their coordinates
$Q_1$, $Q_2$. Following the analysis in \cite{Han97}, they can
be written as two coupled QLE,  
\bean
\ddot Q_1(t) + \omega_0^2 Q_1(t) &+& \frac{d}{dt} \int_0^t
dt'\,\bigl[\Gamma_0(t - t') Q_1(t')\nn \\ 
&+& \Gamma_r(t - t') Q_2(t')\bigr] =  B_1(t) \label{eq:qle1}
\eean
and a similar equation where $1$ and $2$ are interchanged.
Here, we introduced a distance $d$ dependent damping kernel
$
\Gamma_d(t)  = \gamma \Omega(e^{-\Omega\left| t - d \right|} +
e^{-\Omega\left| t + d \right|})\:, 
$
and bath operators
\be
B_{\nicefrac{1}{2}}(t) = \sum_k 
\tilde g_k \cos\fr{kr}{2} e^{i \omega_k t} b_k^\dagger \: \pm \:
\tilde g_k \sin \frac{k r}{2} e^{i
  \omega_k t} a_k^\dagger \:+ \: h.c.\, ,
\ee
where $ \tilde g_k = (\hbar g_k^2 / m_k\omega_k m^2)^{\nicefrac{1}{2}} $,
and $b_k^\dagger,b_k$ and $a_k^\dagger, a_k$ are creation and annihilation operators of
a symmetric and antisymmetric bath mode $k$. Note that the operators
$B_{\nicefrac{1}{2}}(t)$ evolve freely 
in time; the back-action of
the two oscillators on the bath modes is solely contained in the
memory terms in the QLE. The QLE also have a clear classical
interpretation: the two oscillators are subjected to friction with a
damping constant $\gamma$, they are coupled via a bath-mediated
retarded interaction, and they are
exposed to stochastic forces $B_{\nicefrac{1}{2}}(t)$. 
Without $V_c$ the QLE also would exhibit terms proportional to $Q_1(t)
Q_2(t)$, corresponding to  
an instantaneous, direct coupling of the two oscillators. In
principle, the appearance of such term is possible because our model
does not obey Lorentz invariance. Nevertheless, here we are interested
in the bath-mediated coupling of the oscillators, and therefore
eliminated the direct couplings by adding $V_c$ to the system Hamiltonian.  

The formal solution of the QLE is simple, once they are written in
the form 
\ben\label{simple-qle}
\dot{\mathbf y}(t) + \cl Z \mathbf y(t) + \frac{d}{dt}\int_0^t dt'\,
\cl C(t-t')\mathbf y(t') = \mathbf B(t)\:,
\een
where $\b y = (Q_1,Q_2,\dot Q_1, \dot Q_2)$, $\b B = (0,0,B_1,B_2)$,
and $\cl Z$ and $\cl C(t)$ are $4 \times 4$
matrices whose definitions become obvious by comparison of Eq.\
\Ref{simple-qle} with the original QLE. 
Then, the solution $\b y(t)$ of Eq.~\Ref{simple-qle} for initial
$\b y(0)$ and inhomogeneity $\b B(t)$ is 
\ben\label{yoft}
\b y(t) = \cl G(t)  \b y(0)  +  \int_0^t dt' \: \cl G(t-t')\b B(t')\:,
\een
where the Green's function $\cl G(t)$ solves the homogeneous part of
Eq.~\Ref{simple-qle}. Its Laplace transform $\hat{\cl G}(s) = [ s  +
\cl Z + s \hat{\cl C }(s)]^{-1} $ can be calculated analytically.

Correlations and entanglement in the two
oscillator system can be studied on the basis of the
oscillator's dimensionless covariance matrix $C$,
\be
C_{lm} = \sp{\tilde{\b y}_l \tilde{\b y}_m + \tilde{\b y}_m \tilde{\b y}_l }_{\rho_s} 
\equiv \tr[(\tilde{\b y}_l \tilde{\b y}_m + \tilde{\b y}_m \tilde{\b
  y}_l )\rho_s] \:,
\ee
where $\rho_s$ is the joint state of the system oscillators.
The vector $\tilde{\b y}$ is obtained from $\b y$ by multiplying the
first and second entry with $(m\omega_0/\hbar)^{\nicefrac{1}{2}}$, and the
third and forth entry with $(m/\hbar\omega_0)^{- \nicefrac{1}{2}}$.
Assuming that at time $t=0$ the total state factorizes in an
initial oscillator state $\rho_s$ and a thermal state $\rho_T$ of the
bath, the temporal evolution of the covariance matrix follows with
Eq.~\Ref{yoft} to be
\be
C(t) = \cl G(t) C(0) \cl G(t) \dag + \int_0^t \!\!\!\! dt' \!\! \int_0^t \!\!\!\! dt'' \cl
G(t-t') \cl K(t'-t'')\cl G(t-t'')\dag\:.
\ee
Here, $C(0)$ is the covariance matrix of the initial oscillator state
$\rho_s$. The matrix $\cl K(t) = 2m  \sp{ \b B(t) \b
  B(0)^\dagger}_{\rho_T}/\hbar \omega_0 $ contains the correlations of  
the bosonic fields $B_{\nicefrac{1}{2}}$. Its non-vanishing entries are 
$\cl K_{34}(t) = \cl K_{43}(t)$, equal to
\be
\fr{8 \gamma}{\pi \omega_0} \int_0^\infty \!\!\!\!\! d \omega 
\:\omega \fr{\Omega^2}{\Omega^2 + \omega^2} \coth\fr{\omega }{2 T}
\cos \omega t \cos \omega r\:, 
\ee
and two diagonal elements $\cl K_{33}(t) = \cl K_{44}(t)$, which are
given by the same expression, but with $r=0$.

We quantify the amount of entanglement of the
two oscillators by the logarithmic negativity $E$. In
case of a Gaussian state $\rho_s$ of the oscillators, $E$
can be conveniently determined from the correlation matrix $C$ as follows:
First, one applies a time-reversing operation \cite{Sim00} on the second
oscillator, according to which the covariance matrix 
transforms to $C_{lm} \to \tilde C_{lm} = (-1)^{\delta_{l4} +
  \delta_{m4}} C_{lm}$. Then, the symplectic eigenvalues $\lambda_1, \lambda_2$ of $\tilde C$ yield the logarithmic negativity as
$
E = - \sum_{j=1}^2 \log_2 \min(1,\lambda_j)
$ \cite{VW02}.
In this way the entanglement dynamics of the two oscillators 
follows from the temporal evolution of the correlation matrix $C$,
provided that the oscillator state is Gaussian for all times. 
This is the case when we restrict ourself to Gaussian initial states,
since this property is conserved under the dynamics 
of the quadratic Hamiltonian. 
Note that for Gaussian states a vanishing logarithmic
negativity is equivalent to separability \cite{Sim00}.

Having outlined our model and the methods we have used, let us now
present our results. Our main interest is the generation of
entanglement from an initially separable oscillator state $\rho_s(0)$ 
via the coupling with the bosonic bath. Since there are no reasons 
for certain initial separable states being preferred to other
ones, here we present only results where 
initially the system oscillators are in their ground state. 
The bath is assumed to be initially in a thermal state $\rho_T$ of
temperature $T$. Thus, the total state is Gaussian and we
can determine the logarithmic negativity $E$ as a function of
time as outlined above. In the following, we will mainly show numerical data
demonstrating the characteristic dependence of the entanglement
generation on distance $r$, cut-off frequency $\Omega$, damping constant $\gamma$,
and temperature $T$. Generally, we measure distances in units of
$c/\omega_0$, frequencies in units of $\omega_0$, and temperature in
units of $\hbar \omega_0/k_B$.

First, we consider the entanglement of the two oscillators at large
times.
For any finite distance $r$ one finds $\cl G(t) \to  0$ for
$t \to \infty$,
meaning that the initial oscillator state becomes irrelevant at
large times. Hence, the asymptotic covariance matrix is 
$
C_\infty =
 \int_0^\infty \!\! dt' \!\! \int_0^\infty \!\! dt'' \cl
G(t') \cl K(t'-t'')\cl G(t'')\dag\:.
$
The time integrations together with the oscillating factors in $\cl
K(t'-t'')$ represent Laplace transformations,
which eventually result in a single $\omega$ integral over 
terms containing the factor $| \hat{\cl G}(i\omega) |^2$.
The remaining integration over $\omega$ can be easily performed
numerically.
\begin{figure}
  \begin{center}
 \includegraphics{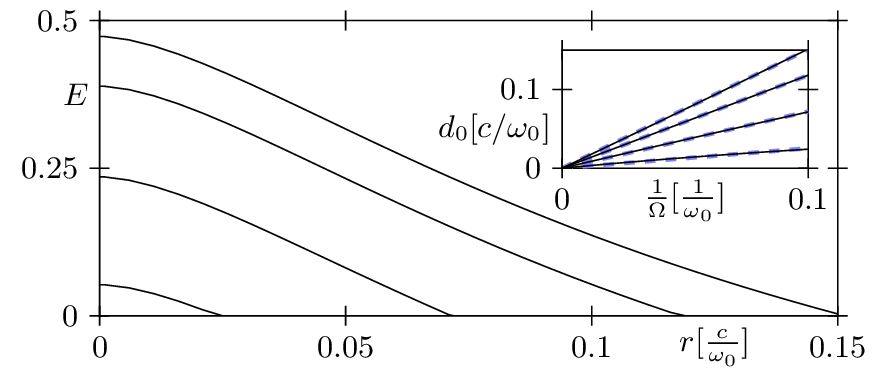}
\caption{\label{asymptotic-distance}
  Asymptotic entanglement of the system oscillators measured in
  logarithmic negativity $E$ as a function of distance $r$ (in units of $c/\omega_0$)
  for temperatures $T = 0$, $0.1$, $0.2$, and $0.3\times \hbar \omega_0 /k_B$
  (upper to lower curves), damping constant $\gamma =\omega_0$, and
  cut-off frequency $\Omega = 10 \omega_0$. 
  $E$ drops to zero at a rather small critical distance $d_0\lesssim
  c/\Omega$, which for the above temperatures is proportional to
  $\Omega\inv$ (cf.~inset). The fitted straight 
  lines (dashed) have slopes $a=1.51$, $1.18$, $0.72$, $0.25$. 
  \vspace{-0.8cm}
}
\end{center}
\end{figure}
Fig.~\ref{asymptotic-distance} shows the asymptotic logarithmic
negativity $E$ as a function of the distance $r$ between the
oscillators. Clearly, the entanglement decreases with distance and
drops to zero at rather small critical distances  $d_0$.
The dependence of $d_0$ on the inverse cut-off frequency 
$\Omega$ for different temperatures can be seen in the inset of
Fig.~\ref{asymptotic-distance}. 
For $\Omega \gtrsim \omega_0$ we find the critical distance $d_0$ to
be inversely proportional to the cut-off frequency, $d_0 \approx a
c/\Omega$, where $a$ is a coefficient of order unity (at $T=0$) that
decreases with increasing temperature. 
The distance $d_0$ is rather insensitive to the actual value of the
damping constant $\gamma$.  
For instance, the critical distance of $d_0=0.151$ (in units of
$c/\omega_0$) at $\gamma = \omega_0$ and $T=0$ just changes to $0.12$
or $0.17$ when the damping is increased or lowered by a factor of 10,
respectively. 

\begin{figure}
\begin{center}
\includegraphics{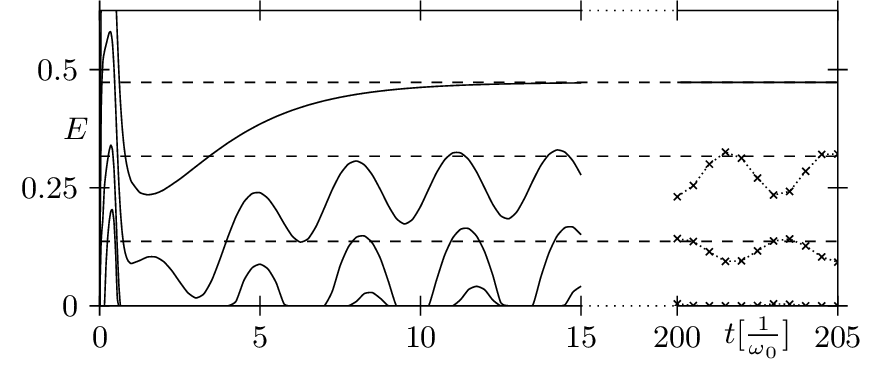}
\caption{\label{dynamical-entanglement}
Logarithmic negativity $E$ as function of time (in units of
$1/\omega_0$) for distances $r=0$, $0.05$, $0.1$, and $0.15\times c/\omega_0$  
(upper to lower curves) below the critical distance $d_0$, $T=0$,
$\Omega=10\omega_0$, and 
$\gamma = \omega_0$. Dashed lines represent asymptotic values.
\vspace{-0.8cm}
}
\end{center}
\end{figure}

Now we consider how the logarithmic negativity
develops in time. Determining the time dependent covariance matrix
$C(t)$ involves an inverse Laplace transformation, which we performed
numerically using Durbin's formula \cite{PH84}.
Results for vanishing and three nonvanishing distances $r$ below the
critical distance $d_0$ are shown in
Fig.~\ref{dynamical-entanglement}. 
All curves show a characteristic peak at short times within which
the logarithmic negativity reaches its maximum value $E_{max}$.
After its decay the logarithmic negativity slowly recovers in an
oscillatory manner to its  asymptotic value, where the frequency of
the oscillation is approximately $\omega_0/2$. The oscillations decay
rather slowly with time because the relative coordinate $Q_1-Q_2$ of
the two oscillators is weakly damped for the small distances $r$ under
consideration.  
This behavior does not change much for distances slightly above $d_0$.
However, at larger distances $r>0.18 c/\omega_0$ the logarithmic
negativity does not recover at all but remains zero for all later times.

\begin{figure}
\begin{center}
\includegraphics{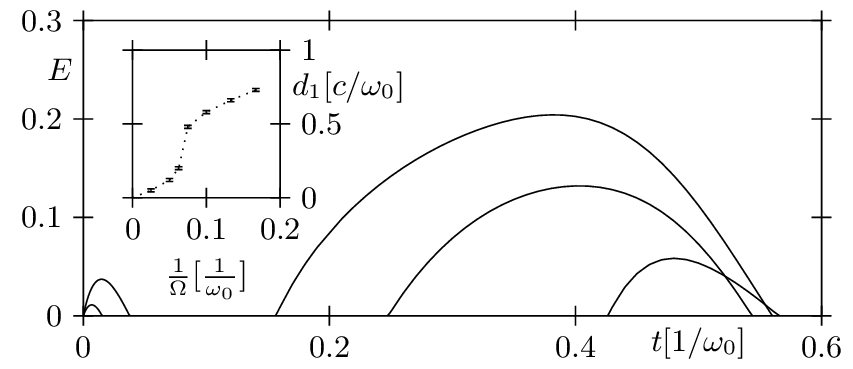}
\caption{
\label{short-time-entanglement}
Short time behavior of $E$ for $r=0.15$, $0.2$, and $0.4$ (in units
of $c/\omega_0$, top to bottom).
The initial peak of $E(t)$ visible in Fig.~\ref{dynamical-entanglement} resolves in two
peaks. The first peak is exponentially suppressed in $2 r
\Omega/c$. The second peak is delayed by approximately $r/c$. Its 
height decreases with $r$ and vanishes for distances $r\ge d_1$. 
The inset shows $d_1$ as function of the inverse cut-off frequency $1/\Omega$
(in units of $1/\omega_0$).
\vspace{-0.8cm}
}
\end{center}
\end{figure}
Focussing on the short time behavior of $E$, 
the initial peak actually resolves in two peaks, 
as shown in Fig.~\ref{short-time-entanglement}.
The first peak appears immediately after switching on the interaction
at times $t$ less than $r/c$. Since bosons cannot have been exchanged
between 
the two system oscillators within this time span, we attribute this peak to
entanglement that had been already present in the bath \cite{SW85}.
Switching on the interaction might immediately transfer part of that
entanglement to the oscillators. 
This behavior can be addressed by a short time expansion of
$C(t)$, which, at zero temperature, eventually results in 
\ben\label{linear-expansion}
E(t) \approx \fr{4}{\ln 2} \:\fr{\gamma}{\omega_0}\left\{
e^{- \fr{r\:\Omega}{c}}\: \Omega t
\:-\:
\alpha_{\Omega t}\: (\Omega t)^2 
\: + \cl O(\Omega t)^3 
\right\}\:,
\een
with an $\Omega t$ dependent 
$\alpha_{\Omega t} \approx 0.2937 -\fr{1}{\pi} \ln \Omega t$.
>From this expansion we find that for $r \gtrsim c/\Omega$ width and height
of the first peak are exponentially suppressed in the parameter
$r\Omega/c$ by factors less than 
$\sim \exp(- r\Omega/c)$ and $\sim \exp(- 2r\Omega/c)$, respectively. 

The second peak in the logarithmic negativity is delayed by
a little bit more than $r/c$, which suggests that it refers
to entanglement due to exchange of bosons. 
Its height decreases monotonically with distance $r$ and, in fact,
reaches zero at a relatively small distance $d_1$ that 
is constrained by the inverse cut-off
frequency. At zero temperature and damping $\gamma = \omega_0$
numerical data show $d_1(\Omega) \lesssim 6.0 c/\Omega$ (cf.~inset of
Fig.~\ref{short-time-entanglement}). 
We expect that the actual value of the damping constant $\gamma$ has
only minor influence on $d_1$ (as like on the distance $d_0$), since 
numerical data as well as Eq.~\Ref{linear-expansion} show that
in first approximation $\gamma$ scales only the amplitude of $E(t)$.

We conclude that generally for distances $r$ significantly larger than
$c/\Omega$ the logarithmic negativity $E(t)$ reaches its total maximum
$E_{max}$ within an exponentially short time $t_0 \lesssim \exp(-r\Omega/c)/\Omega$ and 
then vanishes for all times $t \gtrsim 2 t_0$. Moreover, at these
distances the maximum value $E_{max}$ is exponentially suppressed in
$2 r \Omega/c$. 

To summarize, by analyzing the time-dependent logarithmic negativity 
of two oscillators coupled to a bosonic bath we found strong evidence 
that the entanglement mechanism under consideration is limited 
to rather small distances $r$ of order of $c/\Omega$, i.e. to
distances of order of the cut-off wavelength $\lambda$. In practice, this
length corresponds to the spatial extension of the microscopic systems
to be entangled. 
At larger distances the maximum achievable logarithmic negativity is
exponentially suppressed in, roughly, $r/\lambda$.
We believe that this behavior is characteristic for 
bath-mediated entanglement in general, since there seem to be
no features of the investigated oscillator model which would it
make special for entanglement.  
In fact, the general picture outlined here is fully
supported by results that we obtained for an alternative
two-spin-boson model \cite{QZK09}.
Having said this, one may summarize our findings 
by stating that generally two objects can only be efficiently entangled
via the interaction with a heat bath if they are in immediate vicinity
of each other. 

It might appear puzzling that the environment quickly and strongly
entangles with 
the two oscillators (which, after all, is the
origin of the ubiquitous phenomenon of decoherence), while the two
oscillators for their own remain essentially disentangled (if
they are remote from each other). 
The reason behind this strongly asymmetric behavior
is the large asymmetry in the (effective) Hilbert space
dimensions of the participating systems: 
{\em few} oscillator states interact with a {\em continuum} of bath states.
Assuming that the {\em generic} state of the joint system is well
represented by a randomly chosen state of the joint system, it
follows from \cite{HLW06} that for dimensional reasons the bath is 
strongly entangled with each oscillator, while the system oscillators
on their own remain separable.
Thus, our analysis particularly demonstrates that 
under the actual dynamics
 -- generated by a standard bilinear system-bath interaction --
a non-generic initial state rapidly evolves to a generic one.
Interestingly, the considered interaction fails to produce this effect if 
the distance $r$ becomes less or of the order of the cut-off wavelength
$\lambda$, as evidenced in significant entanglement of the system
oscillators in this case.

We would like to thank C.~Kiefer for valuable discussions. The work is
supported by DFG grant TL2159 and by the Bonn-Cologne Graduate School.


\end{document}